\documentclass[aps,10pt,prb,twocolumn,showpacs,amsmath,amssymb,superscriptaddress]{revtex4-1}

\usepackage{graphicx}
\usepackage{epstopdf}
\usepackage{hyperref}
\usepackage{commath}
\usepackage{dsfont}

\newcommand{\bs}{\boldsymbol}

\begin{document}

\title{The topological Anderson insulator phase in the Kane-Mele model}

\author{Christoph P. Orth\footnote{Correspondence should be addressed to C.P.O.~(email: mail@c-orth.de)}}
\author{Tibor Sekera}
\author{Christoph Bruder}
\affiliation{Department of Physics, University of Basel,
Klingelbergstrasse 82, 4056 Basel, Switzerland}
\author{Thomas L. Schmidt}
\affiliation{Physics and Materials Science Research Unit, University of Luxembourg, L-1511~Luxembourg}
\date{\today}

\begin{abstract}
  It has been proposed that adding disorder to a topologically trivial mercury telluride/cadmium telluride (HgTe/CdTe)
  quantum well can induce a transition to a topologically nontrivial
  state. The resulting state was termed topological Anderson insulator and was
  found in computer simulations of the Bernevig-Hughes-Zhang model.
  Here, we show that the topological Anderson insulator is a more
  universal phenomenon and also appears in the Kane-Mele model of topological insulators on a honeycomb lattice. We
  numerically investigate the interplay of the relevant parameters,
  and establish the parameter range in which the topological Anderson
  insulator exists.
  A staggered sublattice potential turns out to be a
  necessary condition for the transition to the topological Anderson
  insulator. For weak enough disorder, a calculation based on the lowest-order Born approximation reproduces quantitatively the numerical data.
  Our results thus considerably increase the number of candidate materials for the topological Anderson insulator phase.
\end{abstract}

\pacs{73.20.Fz,03.65.Vf,73.43.Nq}
\maketitle

Topological insulators (TIs) are novel materials which have raised a
great deal of interest over the past decade.\cite{hasan10, qi11} One
of their distinguishing features is the existence of
conducting boundary states together with an
insulating bulk. The boundary states are protected by time-reversal symmetry (TRS) and exist both in two-dimensional (2D) and
three-dimensional (3D) TIs. In 2D TIs, the boundary states lead to an
edge conductance of one conductance quantum per edge for chemical potentials inside the bulk band gap.\cite{bernevig06, kane05, kane05b}

It is a challenging task to find candidate materials for TIs. So far,
only a limited number of materials are known. The most prominent 2D TIs
are HgTe/(Hg,Cd)Te quantum wells (HgTeQWs)~\cite{koenig07} and InAs/GaSb heterostructures,\cite{knez11,li15} whereas 3D
TIs were found for instance in Bi$_{1-x}$Sb$_x$.
\cite{hsieh08} The fact that their metallic surface states emerge due to a topological property of the bulk band structure means that they are robust to weak disorder. However, one expects that a large amount of disorder should ultimately localize the surface states and render them insulating.

All the more surprising, it was predicted that
the opposite transition can happen in certain parameter ranges: adding strong disorder can convert a trivial insulator without edge states into a topological insulator with perfectly conducting edge states. Materials that exhibit this new state have been
termed topological Anderson insulators (TAIs).

This effect was first theoretically predicted based on the lattice version of the Bernevig-Hughes-Zhang (BHZ)
model for HgTeQWs in the presence of Anderson disorder.\cite{Li09,jiang09} For Anderson disorder, a random
on-site potential, uniformly distributed in an energy window of width $2W$, is assigned to each lattice site of a tight-binding model. From Anderson's theory of localization~\cite{anderson58} one expects that
a system with finite conductance without disorder undergoes a transition to a system with localized states and suppressed conductance as the disorder is increased beyond a certain threshold value. The behavior of TAIs instead is quite different. A TAI is an ordinary band insulator in the clean limit. Above a critical disorder strength $W$, an interesting topological state appears, in which the material features a quantized conductance. For even stronger $W$, above the disorder strength at which the states of the conduction and valence band localize, it was proposed that tunneling across the bulk becomes possible,\cite{chen12} probably enabled by percolating states,\cite{girschik15} and the conductance is again suppressed.

The disorder-induced transition can be understood by a renormalization
of the model parameters. The BHZ model with disorder and band
mass $m$ can be approximated by an effective model of a clean system
and renormalized mass $\bar m$. Using an effective-medium
theory and the self-consistent Born approximation (SCBA), it was shown that for
certain model parameters, $\bar m$ can become
negative even if the bare mass $m$
is positive.\cite{groth09} As a consequence, the effective model becomes that of a
TI and features edge states with a quantized conductance of
$G_0=e^2/h$.\cite{prodan11}

Furthermore, TAIs have been predicted in several related systems, for instance in a honeycomb lattice described by the time-reversal-symmetry breaking Haldane model,\cite{xing11} a modified Dirac model,\cite{xing11} the BHZ model with $s_z$ non-conserving spin-orbit coupling,\cite{yamakage11} as well as in 3D topological insulators.\cite{guo10} Moreover, similar transitions from a topologically trivial to a topologically nontrivial phase have been found to be generated by periodically varying potentials~\cite{fu14} or phonons.\cite{garate13} In contrast to on-site Anderson disorder, certain kinds of bond disorder cannot produce a TAI as they lead only to a positive correction to $m$.\cite{song12, lv13} So far, however, the TAI phase was not found in the Kane-Mele model on a honeycomb lattice, describing for example graphene or proposed TIs such as silicene, germanene and stanene.\cite{aufray10, kara12, davila14, zhu15} First indications to this phase were already found, showing that the Kane-Mele model without a staggered sublattice potential hosts extended bulk states even for large disorder strengths.\cite{prodan11b}

In this paper we show the existence of TAIs in the Kane-Mele model by
means of tight-binding calculations. The interplay between the parameters characterizing intrinsic spin-orbit coupling (SOC)
$\lambda_\text{SO}$, extrinsic Rashba SOC $\lambda_R$, and a staggered
sublattice potential $\lambda_\nu$ turns out to be crucial for the
visibility of TAIs, and we calculate the parameter ranges
in which TAIs can be observed. We find analytically that
to lowest order in $W$, the parameters $\lambda_\text{SO}$ and
$\lambda_R$ are not renormalized with increasing disorder strength, in contrast to $\lambda_\nu$. However, a new effective hopping $\lambda_{R3}$ is generated due to the disorder, which is related but not identical to $\lambda_R$. Although $\lambda_R$ is not a crucial ingredient for the existence of TAIs, it significantly alters the physics of topological insulators in various ways \cite{orth13, rod15} and, as we will show below, strongly affects the TAI state.

Even though recently first signs of a TAI phase may have been found experimentally in evanescently coupled waveguides,\cite{stutzer15} there has been no experimental evidence so far for the existence of the TAI phase in fermionic systems. The main difficulty is the requirement of a rather large and specific amount of disorder, which is difficult to control in the topological insulators currently investigated, where the 2D TI layer is buried inside a semiconductor structure. In contrast, producing and controlling disorder in 2D materials described by the Kane-Mele model could be much easier. Disorder in 2D materials with honeycomb structure can be produced by randomly placed adatoms \cite{weeks11, hua12} or a judicious choice of substrate material.\cite{ando06, ishigami07, fratini08, varlet15} Moreover, a sizeable staggered sublattice potential can be generated via a suitable substrate material.\cite{nevius15} Other means of engineering disorder were proposed in periodically driven systems.\cite{titum15, yang15} Finally, honeycomb structures with the SOC necessary to produce a topological phase have already been realized using ultracold atoms in optical lattices,\cite{jotzu14} in which disorder can in principle be engineered.

\section*{Results}
\subsection*{Setup}
The basis of our calculations is the Kane-Mele model~\cite{kane05b} given by the following Hamiltonian on a tight-binding honeycomb lattice
\begin{align} \label{eq:KaneMeleHamiltonian}
	H &=
 t \sum_{\langle i j \rangle} c_i^\dag c_j + i \lambda_\text{SO} \sum_{\langle \langle i j \rangle \rangle} \nu_{ij} c_i^\dag s^z c_j + \lambda_\nu \sum_i \xi_i c_i^\dag c_i\notag \\
&+ i \lambda_R \sum_{\langle i j \rangle} c_i^\dag \left( {\bf s} \times {\bf \hat d}_{ij}\right)_z c_j
 + W \sum_i \epsilon_i c_i^\dag c_i\:,
\end{align}
which has been supplemented by an on-site Anderson disorder term with disorder
strength $W$ and uniformly distributed random variables $\epsilon_i
\in [-1,1]$. The summations over the lattice sites $\langle ij \rangle$ and  $\langle \langle
i,j \rangle \rangle$ include all nearest neighbors and next-nearest
neighbors, respectively. The operators $c_i^\dag = (c_{i\uparrow}^\dag,c_{i\downarrow}^\dag)$,
$c_i=(c_{i\uparrow}, c_{i\downarrow})^T$ are creation and annihilation operators for the site $i$ of the lattice. The parameters $t$, $\lambda_\text{SO}$ and $\lambda_R$ are the nearest-neighbor hopping strength, intrinsic SOC, and Rashba SOC, respectively. If the next-nearest neighbor hopping from site $j$ to site $i$ corresponds to a right-turn on the honeycomb lattice, then $\nu_{ij} = 1$, otherwise $\nu_{ij} = -1$. Furthermore, $\bs s = (s^x, s^y, s^z)$ is the vector of Pauli matrices for the spin degree of freedom, and ${\bf \hat d}_{ij}$
is the unit vector between sites $j$ and $i$.
The Wannier states at the two basis atoms of the honeycomb lattice are separated in energy by twice the staggered sublattice potential $\lambda_\nu$, with $\xi_i=1$ for the $A$ sublattice and $\xi_i=-1$ for the $B$ sublattice. The lattice constant is $a$.
\begin{figure*}
  \includegraphics{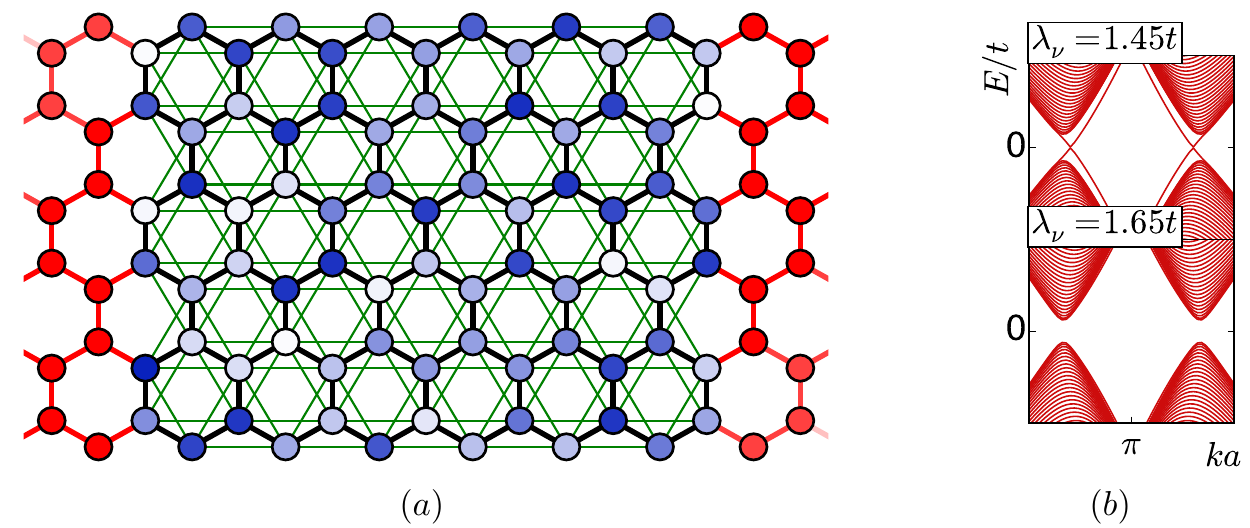}
\caption{(a) Toy model illustrating the
  tight-binding terms in the Kane-Mele Hamiltonian~\eqref{eq:KaneMeleHamiltonian}. The blue color scale marks different
  on-site potentials. Thick black lines correspond to nearest-neighbor
  hopping and Rashba SOC, while thin green lines correspond to
  intrinsic SOC. The leads attached at both sides (red color) are
  modeled by a hexagonal lattice with nearest-neighbor hopping term
  and finite chemical potential. In this example, the sample has width
$w=5a$ and length $l=6a$.  Much larger sample sizes of $w=93a$ and
length $l=150a$ were used in the calculations described below. (b) Band
structures of infinitely long samples of width $w=93a$ for two different
values of $\lambda_\nu$ showing a topologically nontrivial and a trivial gap.
Vertical and horizontal axis correspond to energy in units of $t$ and
dimensionless momentum, respectively. Parameters are
$\lambda_\text{SO}=0.3t$ and $\lambda_R=0$.}
\label{fig:small_sample}
\end{figure*}

The band structure of this model depends strongly on the parameter set
$\lambda_\text{SO}$, $\lambda_R$, and $\lambda_\nu$. In the clean limit and for $\lambda_R = 0$, the system will be a topological insulator for $\lambda_\nu/\lambda_\text{SO} < 3 \sqrt{3}$ and a trivial insulator otherwise.\cite{kane05b} The tight-binding lattice and examples for the band structure in the clean limit are displayed in Fig.~\ref{fig:small_sample}. For $W = \lambda_\nu = 0$, the system will be a topological insulator if $\lambda_R/\lambda_\text{SO} \lesssim 2 \sqrt{3}$ and a metal or semimetal otherwise. For finite $\lambda_\nu$ and $\lambda_R$ the situation is more complex and a topological transition appears for values within these two boundaries.

\begin{figure}[t]
\centering
\includegraphics[width=8.5cm]{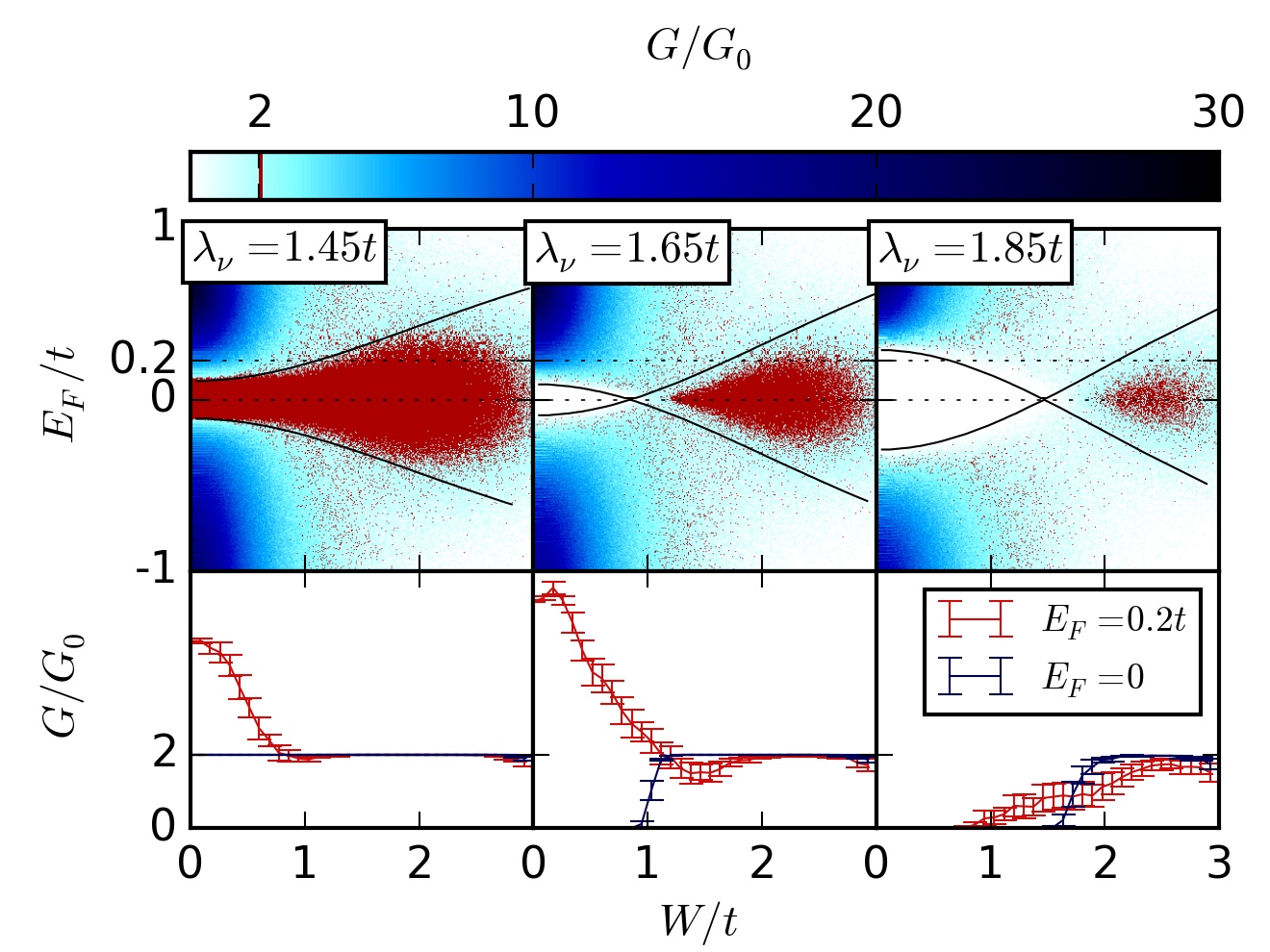}
\caption{\emph{Top row:} The conductance from the left to the right
  lead as a function of the disorder strength $W$ (horizontal axis)
  and chemical potential $E_F$ (vertical axis). The conductance varies from $0$
  (white) to $30 G_0$ (dark blue). The quantized value of $2 G_0$
  (red for all conductances within $[1.95 G_0, 2.05 G_0]$) originates from two helical edge states. The three plots show
  the conductance for three different values of $\lambda_\nu$
  that represent, respectively, a topological insulator, a TAI, and a TAI at the
  transition to an ordinary insulator. The black lines are obtained
  from a lowest-order Born approximation without any adjustable
  parameter. The two dotted lines mark the energies $E_F = 0$,
  $E_F=0.2t$.
  \emph{Bottom row:} The conductance at fixed chemical potentials $E_F=0$
  (black) and $E_F=0.2 t$
  (red) for the same parameters as in the top row. The errors bars
  originate from an averaging procedure over 100 disorder
  configurations. The vanishing error bars in the regions with
  a conductance of $2G_0$ show that the topological phase is stable irrespective
  of the exact disorder configuration. The system parameters are $w=93 a$,
  $l=150 a$, $\lambda_\text{SO}=0.3 t$, and $\lambda_R=0$.}
  \label{fig:heatmaps_lambdaNusweep}
\end{figure}

\subsection*{Numerical solution}
For $\lambda_R=0$, we find a TAI phase for parameters close to the
topological transition at $\lambda_{\nu}/\lambda_\text{SO} = 3
\sqrt{3} \approx 5.2$. Changing this ratio corresponds to changing the band mass in the
case of the BHZ model.
Figure~\ref{fig:heatmaps_lambdaNusweep} shows the conductance for
different values of $\lambda_\nu$. We find that for $\lambda_\nu =
1.45 t \approx 4.8 \lambda_\text{SO}$ the system is a topological insulator. For $W=0$, i.e., in the
clean case, the conduction and valence bands are separated by a red
region with a quantized conductance of $2 G_0$. Remarkably, with
increasing disorder strength, the states in the conduction and valence
bands localize, but the helical edge states that are responsible for
the conductance of $2G_0$ exist for an even larger energy window. The
conductances and the vanishing error bars for the two distinct energy
values $E_F = 0$, $E_F = 0.2 t$ in the lower row of
Fig.~\ref{fig:heatmaps_lambdaNusweep} show that the conductance
quantization, and with it the topological nature of the system,
persist for the vast majority of microscopic disorder
configurations. Interestingly, for $\lambda_\nu = 1.65t = 5.5 \lambda_\text{SO}$, the system is
a trivial insulator at $W=0$. The trivial gap closes however, and at
$W\approx t$
the system develops a topologically non-trivial gap and edge
states. This can be seen from the quantized conductance. Finally, for
$\lambda_\nu=1.85 t \approx 6.2 \lambda_\text{SO}$, the closing of the trivial gap and re-opening of the topological gap happens at a disorder strength which is strong enough to destabilize the emergent topological phase. Features of the conductance quantization can still be seen, but this behavior is not that robust anymore. As no averaging is done in the upper row of Fig.~\ref{fig:heatmaps_lambdaNusweep}, and a new disorder configuration is taken for every data point, destabilization of the topological phase can be seen by red and white speckles in the figure.

\begin{figure}[t]
\centering
\includegraphics[width=8.5cm]{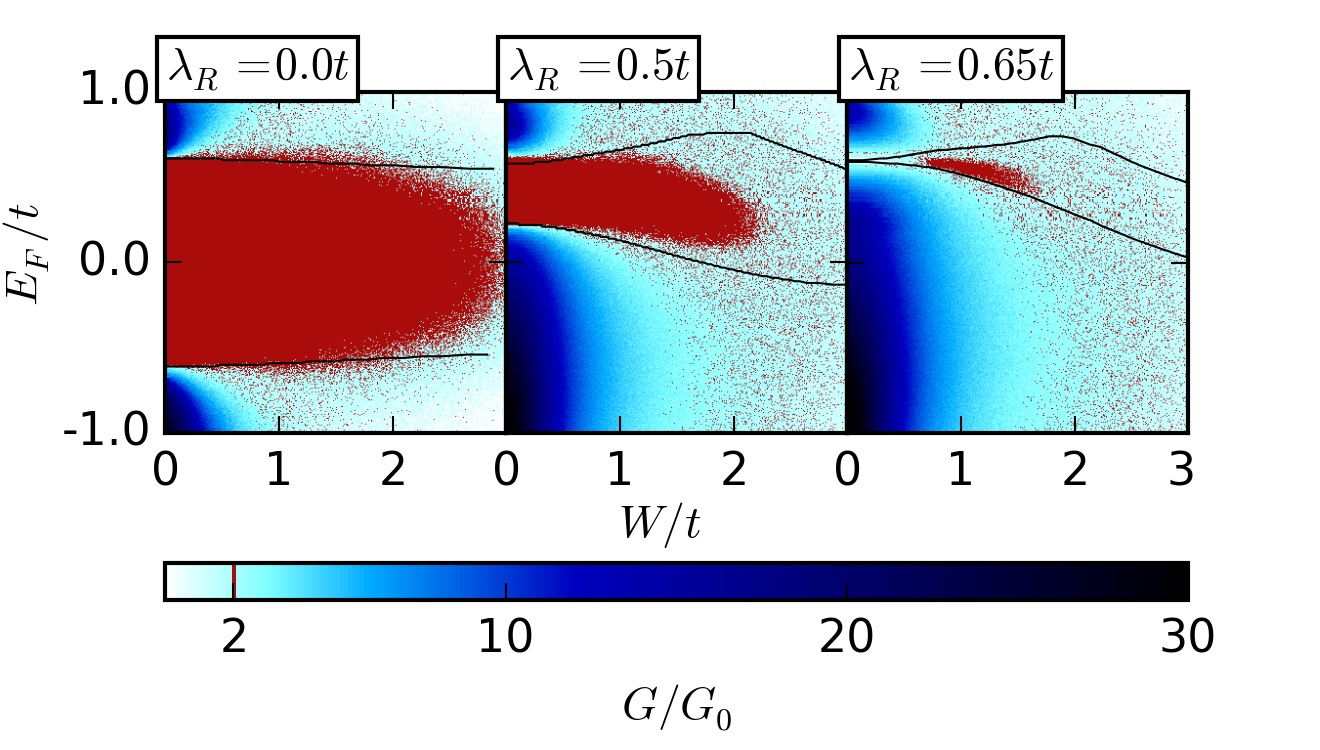}
\caption{The conductance for increasing disorder strength $W$ and
  chemical potentials $E_F$ for three different values of $\lambda_R=0$
  (left), $\lambda_R=0.5 t$ (middle) and $\lambda_R=0.65 t$
  (right). The system parameters are $w=93 a$, $l=150 a$, $\lambda_\text{SO}=0.3 t$, and $\lambda_\nu=0.95 t$. The black lines are obtained from a lowest-order Born approximation without any fitting parameter. The conductance color code is the same as in Fig.~\ref{fig:heatmaps_lambdaNusweep}}
\label{fig:heatmaps_lambdaRsweep}
\end{figure}

We find that no TAI exists without staggered sublattice potential ($\lambda_\nu = 0$). If both
$\lambda_\nu$ and $\lambda_R$ are finite, the TAI phase is in general
less pronounced,
see Fig.~\ref{fig:heatmaps_lambdaRsweep}. The plot on the right shows the
closing of a trivial gap and emergence of a topological phase at $W
\approx 0.5 t$.

Furthermore, we observe that the simultaneous presence of intrinsic and Rashba SOC (both
$\lambda_R \neq 0$ and $\lambda_{SO} \neq 0$) destroys
the particle-hole symmetry in the spectrum. In the absence of Rashba SOC, the symmetry operator $\Upsilon$, which acts on the lattice operators as $\Upsilon c_{i\sigma A} \Upsilon^{-1} = c_{-i,\sigma,B}^\dag$ and $\Upsilon c_{i\sigma B} \Upsilon^{-1} = -c_{-i,\sigma A}^\dag$ for the sublattices $A$ and $B$, leaves the (disorder-free) Hamiltonian invariant. $\Upsilon$ can be viewed as particle-hole conjugation combined with spatial inversion, and the inversion is needed to leave the staggered sublattice potential term invariant.

\subsection*{Lowest-order Born approximation}
In the self-consistent Born approximation, the self-energy $\Sigma$ for a finite disorder strength is given by the following integral equation~\cite{bruus04, groth09}
\begin{align} \label{eq:SCBAequation}
\Sigma = \frac{1}{3} W^2 \left(\frac{a}{2 \pi}\right)^2 \int_{BZ} d\bs k \frac{1}{E_F - \mathcal H(\bs k) - \Sigma}\:,
\end{align}
where $\mathcal H(\bs k)$ is the Fourier transform of $H$ in the clean
limit.\cite{kane05b} The coefficient $1/3$ originates from the second
moment $\langle \epsilon_i^2 \rangle=1/3$ of the uniform distribution function
of the disorder amplitudes, and $E_F$ is the chemical potential. The
integration is over the full first Brillouin zone. We use the
lowest-order Born approximation, which means setting $\Sigma=0$ on the
right-hand side of the equation.

After a low-energy expansion of $\mathcal H(\bs k)$, the integral can
be evaluated analytically~\cite{groth09} for $\lambda_R = 0$. This
requires keeping the terms up to second order in $\bs k$ wherever this is the leading $\bs k$-dependent order. The evaluation yields the renormalized staggered sublattice potential
\begin{align}
\bar \lambda_\nu
&=
\lambda_\nu + \frac{W^2}{9 \pi \sqrt{3} \lambda_\text{SO}} \log\abs{\frac{27 \lambda_\text{SO}^2}{E_F^2 - (\lambda_\nu - 3 \sqrt{3} \lambda_\text{SO})^2} \left(\frac{\pi}{2}\right)^4}\:,
\end{align}
For a certain set of parameters, the logarithm can be negative and $\bar \lambda_\nu$ is reduced compared to $\lambda_\nu$. Moreover, we find that $\lambda_\text{SO}$ is not renormalized to order $W^2$. Therefore, it is possible to obtain $\lambda_\nu >3 \sqrt{3} \lambda_\text{SO} > \bar\lambda_\nu$. The system thus makes a transition from a trivial insulator to a topological insulator with increasing $W$.

For a more quantitative treatment, we evaluate the integral for the full Hamiltonian $\mathcal H(\bs k)$ numerically. The self-energy $\Sigma$ is then written as a linear combination of several independent contributions
\begin{align}
  \Sigma = \sum_{a=0}^5 g_a \Gamma^a + \sum_{a<b=1}^5 g_{ab} \Gamma^{ab} ,
\end{align}
with $\Gamma^{(0,1,2,3,4,5)} = (\mathds{1} \otimes \mathds{1},
\sigma^x \otimes \mathds{1}, \sigma^z \otimes \mathds{1}, \sigma^y
\otimes s^x, \sigma^y \otimes s^y, \sigma^y \otimes s^z)$ and
$\Gamma^{ab} = [\Gamma^a, \Gamma^b]/(2i)$. Here, $\sigma^x,
\sigma^y, \sigma^z$ denote the Pauli matrices for the sublattice index. This leads to the following equations for the renormalized quantities
\begin{align}
\bar \lambda_\nu &= \lambda_\nu + g_2 \notag \\
\bar E_F &= E_F - g_0 \notag \\
\bar \lambda_{R3} &= g_3\:,
\end{align}
whereas $\bar \lambda_\text{SO} = \lambda_\text{SO}$ and $\bar \lambda_R =
\lambda_R$ remain unrenormalized to lowest order in $W$. Surprisingly, a new coupling $\bar \lambda_{R3} \Gamma^3$ is created by the disorder. This coupling has the matrix structure $\Gamma^3$, which is similar but not identical to the one for Rashba SOC. Expressing this new term in the lattice coordinates of Eq.~(\ref{eq:KaneMeleHamiltonian}) reveals that it corresponds to a Rashba-type nearest-neighbor hopping term which is asymmetric and appears only for bonds that are parallel to the unit vector $(0,1)$,
\begin{align}
	H_{R3} = i \lambda_{R3} \sum_{\langle i j \rangle_v} c_i^\dag \left( {\bf s} \times {\bf \hat d}_{ij}\right)_z c_j\:,
\end{align}
where $\langle i j \rangle_v$ stands for summations over strictly vertical bonds only. Furthermore, we find to lowest order in $W$ that $\bar\lambda_{R3}=0$ for $\lambda_R=0$.

For $W=\lambda_R=0$, the upper and lower edge of the gap are at the
energies $E_F = \pm |3 \sqrt{3} \lambda_\text{SO}-\lambda_\nu|$. This
is the case for both topological and trivial insulators. Extrapolation
of these equations to finite $W$ leads to the conditions $\bar E_F(E_F) =
\pm |3 \sqrt{3} \bar \lambda_\text{SO} - \bar \lambda_\nu(E_F)|$. The solid
black lines in Fig.~\ref{fig:heatmaps_lambdaNusweep} are the two solutions
to these equations and describe the closing and reopening of the gap
qualitatively for small $W$.

For finite $\lambda_R$ and therefore finite $\lambda_{R3}$, there is no analytical
expression of the gap energy. In this case, we read off the positions of the gap edges
from band structure calculations for several values of $\lambda_R$ and $\lambda_{R3}$. An interpolation
leads to two functions $h_{U,L}(\lambda_\nu, \lambda_R, \lambda_{R3})$ for the upper
and lower band edge in the clean system. Replacing the unperturbed by the renormalized parameters yields two equations
\begin{align}
    h_{U,L}[\bar\lambda_\nu(E_F), \bar\lambda_R(E_F), \bar\lambda_{R3}(E_F)]=\bar E_F(E_F)
\end{align}
The solutions of these equations are indicated by the solid black lines in Fig.~\ref{fig:heatmaps_lambdaRsweep}. Hence, these results agree with the numerical data for small $W$ without any fitting parameter. Deviations appear for larger $W$, when the lowest-order Born approximation is not applicable.

\begin{figure}[t]
\centering
\includegraphics[width=8.5cm]{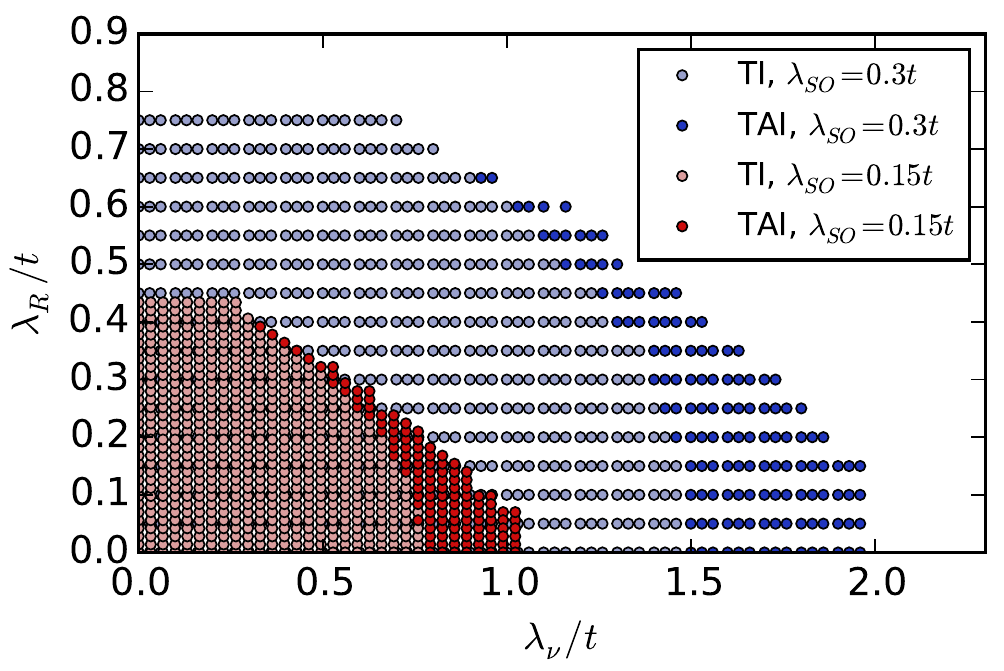}
\caption{Phase diagram in the ($\lambda_\nu$, $\lambda_R$)
  plane. Strong blue (red) color marks the region for which a TAI
  exists for $\lambda_\text{SO}=0.3t$ $(0.15t)$. Transparent blue
  (red) color indicates the regions where a topological insulator is
  found for zero disorder. Each dot represents an individual
  simulation of the kind illustrated in
  Fig.~\ref{fig:heatmaps_lambdaNusweep}.}
\label{fig:phase_diag}
\end{figure}

\subsection*{Phase diagram}
\begin{figure}[t]
\centering
\includegraphics[width=8.5cm]{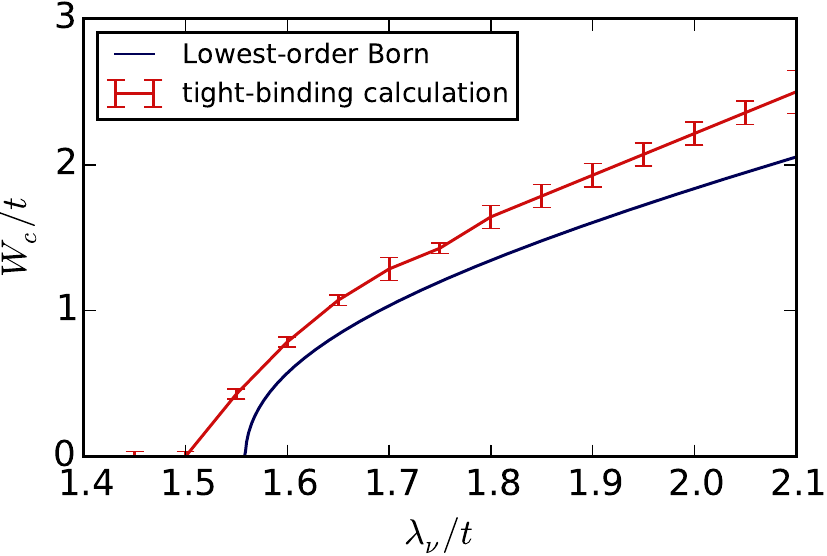}
\caption{Critical value of the disorder strength for the TAI transition along the line
$\lambda_R=0$ for $\lambda_\text{SO}=0.3t$. Comparison between tight-binding simulation and analytical results.}
\label{fig:critical_W}
\end{figure}

Figure~\ref{fig:phase_diag} shows a phase diagram as a function of $\lambda_\nu$ and $\lambda_R$ based on the tight-binding simulations. The dark color marks the regions for
which a critical disorder strength $W_c$ exists above which the system is a TAI (blue for
$\lambda_\text{SO}=0.3t$, red for $\lambda_\text{SO}=0.15t$).
The TAI phase is located
along the boundary separating trivial from
topological insulators in the clean case. Towards larger $\lambda_R$, the TAI region becomes narrower and eventually vanishes above a critical $\lambda_R$. Figure~\ref{fig:critical_W} shows the critical disorder strength $W_c$ as a function of $\lambda_\nu$ for a fixed value of $\lambda_{R}$.

In Figs.~\ref{fig:heatmaps_lambdaNusweep}~and~\ref{fig:heatmaps_lambdaRsweep}
rather large values of the parameters $\lambda_\text{SO},$
$\lambda_\nu$ and $\lambda_R$ were chosen to better visualize the
effect. The TAI phenomenon scales down also to smaller values of the
parameters, as the red region in the Fig.~\ref{fig:phase_diag}
indicates, but the TAI phase becomes less pronounced in the
conductance plots and is harder to identify. Material parameters for
stanene for example are $t=1.3 \text{eV}$,
$\lambda_\text{SO}=0.1 \text{eV}$\cite{xu13} and
$\lambda_R=10 \text{meV}$.\cite{liu11c} We suspect that disorder,
e.g., originating from missing or dislocated atoms, can reach disorder
strengths in the eV range.

\subsection*{Alternative disorder models}

Anderson disorder is a special model for disorder which is not necessarily representative for all TI materials. To better understand the effect of the disorder model, we briefly remark on the following disorder Hamiltonian
\begin{align}\label{eq:Hprime}
  H' &= \frac{W}{\sqrt{3}} \sum_i \eta_i c_i^\dag c_i\:.
\end{align}
In contrast to the Anderson disorder model, where a random potential is assigned to \emph{every} lattice site, here the distribution function for $\eta_i$ is such that only a fraction $0 < \rho \leq 1$ of the sites are affected by disorder. Denoting the total number of sites by $N$, we assume $\eta_i= 1$ on $\rho N/2$ sites, $\eta_i= -1$ on $\rho N/2$ sites, and $\eta_i = 0$ on the remaining sites. The disorder amplitude $W$ is constant. Because $\langle \eta_i^2 \rangle = \rho$, the normalization factor in Eq.~(\ref{eq:Hprime}) ensures that the mean squared disorder strength is equal to the Anderson disorder case for $\rho = 1$.

For general $\rho$, the prefactor in Eq.~(\ref{eq:SCBAequation})
is thus replaced by $\rho W^2/3$. The lowest-order Born approximation for the disorder model (\ref{eq:Hprime}) therefore predicts that a reduced disorder density $\rho$ can be exactly compensated by an increased amplitude $W$. For large enough $\rho$, this is indeed confirmed in the tight-binding simulations.

However, because a single impurity ($\rho = 1/N$) cannot destroy the topological phase, it is clear that the TAI phase should eventually vanish for $\rho \to 0$ at arbitrary $W$. Nevertheless, we find numerical evidence for the TAI phase at surprisingly low impurity densities. A TAI region remains pronounced for densities as low as $\rho = 0.1$.

\section*{Discussion}
In conclusion, we have shown that the topological Anderson insulator
is a significantly more universal phenomenon than previously
thought. Using a combination of an analytical approach and
tight-binding simulations, we have established that the topological
Anderson insulator appears in the Kane-Mele model that describes
potential topological insulators such as silicene, germanene, and
stanene and that can also be realized in optical lattices.  We have observed a transition from a trivially insulating
phase to a topological phase at a finite disorder strength and have
mapped out the phase diagram as a function of the staggered sublattice
potential ($\sim\lambda_\nu$) and the Rashba spin-orbit coupling
($\sim\lambda_R$). The new Anderson insulator exists at the
boundary between trivial and topological insulators for small $\lambda_R$
and finite $\lambda_\nu$,
but not at the boundary between a semimetal and a topological
insulator for small $\lambda_\nu$ and finite $\lambda_R$.
Since the Kane-Mele model on a honeycomb lattice describes a wide
class of candidate materials for topological insulators, we hope that
our work will trigger experimental efforts to confirm the existence of
the topological Anderson insulator.

\section*{Methods}

The numerical simulations were done with the tight-binding Hamiltonian (\ref{eq:KaneMeleHamiltonian}) on a honeycomb lattice with rectangular shape of width $w=93a$ and length $l=150a$ using the Kwant code.\cite{groth14} A smaller version of the sample is shown in Fig.~\ref{fig:small_sample}. Both the upper and lower edge are taken to be of zigzag type. At the left and right edges two semi-infinite, metallic leads of width $w$ are attached. The leads are also modeled by a honeycomb lattice with only nearest-neighbor hopping and a finite on-site energy of $1.2 t$ to bring them into the metallic regime.

\section*{Acknowledgments}
CPO, TS, and CB acknowledge financial support by the Swiss SNF and the
NCCR Quantum Science and Technology.  TLS acknowledges support by
National Research Fund, Luxembourg (ATTRACT 7556175).

\section*{Author contributions}

CPO and TS performed the numerical simulations with input from CB and TLS. All authors contributed to writing the manuscript.

\section*{Additional information}

Competing financial interests: The authors declare no competing financial interests.

\bibliographystyle{naturemag}

\begin{thebibliography}{10}
\expandafter\ifx\csname url\endcsname\relax
  \def\url#1{\texttt{#1}}\fi
\expandafter\ifx\csname urlprefix\endcsname\relax\def\urlprefix{URL }\fi
\providecommand{\bibinfo}[2]{#2}
\providecommand{\eprint}[2][]{\url{#2}}

\bibitem{hasan10}
\bibinfo{author}{Hasan, M.~Z.} \& \bibinfo{author}{Kane, C.~L.}
\newblock \bibinfo{title}{Colloquium: topological insulators}.
\newblock \emph{\bibinfo{journal}{Rev. Mod. Phys.}}
  \textbf{\bibinfo{volume}{82}}, \bibinfo{pages}{3045} (\bibinfo{year}{2010}).

\bibitem{qi11}
\bibinfo{author}{Qi, X.} \& \bibinfo{author}{Zhang, S.}
\newblock \bibinfo{title}{Topological insulators and superconductors}.
\newblock \emph{\bibinfo{journal}{Rev. Mod. Phys.}}
  \textbf{\bibinfo{volume}{83}}, \bibinfo{pages}{1057} (\bibinfo{year}{2011}).

\bibitem{bernevig06}
\bibinfo{author}{Bernevig, B.~A.}, \bibinfo{author}{Hughes, T.~L.} \&
  \bibinfo{author}{Zhang, S.-C.}
\newblock \bibinfo{title}{Quantum spin {Hall} effect and topological phase
  transition in {HgTe} quantum wells}.
\newblock \emph{\bibinfo{journal}{Science}} \textbf{\bibinfo{volume}{314}},
  \bibinfo{pages}{1757} (\bibinfo{year}{2006}).

\bibitem{kane05}
\bibinfo{author}{Kane, C.~L.} \& \bibinfo{author}{Mele, E.~J.}
\newblock \bibinfo{title}{Quantum spin {H}all effect in graphene}.
\newblock \emph{\bibinfo{journal}{Phys. Rev. Lett.}}
  \textbf{\bibinfo{volume}{95}}, \bibinfo{pages}{226801}
  (\bibinfo{year}{2005}).

\bibitem{kane05b}
\bibinfo{author}{Kane, C.~L.} \& \bibinfo{author}{Mele, E.~J.}
\newblock \bibinfo{title}{{$Z_2$} topological order and the quantum spin {H}all
  effect}.
\newblock \emph{\bibinfo{journal}{Phys. Rev. Lett.}}
  \textbf{\bibinfo{volume}{95}}, \bibinfo{pages}{146802}
  (\bibinfo{year}{2005}).

\bibitem{koenig07}
\bibinfo{author}{K\"{o}nig, M.} \emph{et~al.}
\newblock \bibinfo{title}{Quantum spin {Hall} insulator state in {HgTe} quantum
  wells}.
\newblock \emph{\bibinfo{journal}{Science}} \textbf{\bibinfo{volume}{318}},
  \bibinfo{pages}{766} (\bibinfo{year}{2007}).

\bibitem{knez11}
\bibinfo{author}{Knez, I.}, \bibinfo{author}{Du, R.-R.} \&
  \bibinfo{author}{Sullivan, G.}
\newblock \bibinfo{title}{Evidence for helical edge modes in inverted
  {InAs/GaSb} quantum wells}.
\newblock \emph{\bibinfo{journal}{Phys. Rev. Lett.}}
  \textbf{\bibinfo{volume}{107}}, \bibinfo{pages}{136603}
  (\bibinfo{year}{2011}).

\bibitem{li15}
\bibinfo{author}{Li, T.} \emph{et~al.}
\newblock \bibinfo{title}{Observation of a helical {L}uttinger liquid in
  {InAs}/{GaSb} quantum spin {H}all edges}.
\newblock \emph{\bibinfo{journal}{Phys. Rev. Lett.}}
  \textbf{\bibinfo{volume}{115}}, \bibinfo{pages}{136804}
  (\bibinfo{year}{2015}).

\bibitem{hsieh08}
\bibinfo{author}{Hsieh, D.} \emph{et~al.}
\newblock \bibinfo{title}{A topological {Dirac} insulator in a quantum spin
  {Hall} phase}.
\newblock \emph{\bibinfo{journal}{Nature}} \textbf{\bibinfo{volume}{452}},
  \bibinfo{pages}{970} (\bibinfo{year}{2008}).

\bibitem{Li09}
\bibinfo{author}{Li, J.}, \bibinfo{author}{Chu, R.-L.}, \bibinfo{author}{Jain,
  J.~K.} \& \bibinfo{author}{Shen, S.-Q.}
\newblock \bibinfo{title}{Topological {Anderson} insulator}.
\newblock \emph{\bibinfo{journal}{Phys. Rev. Lett.}}
  \textbf{\bibinfo{volume}{102}}, \bibinfo{pages}{136806}
  (\bibinfo{year}{2009}).

\bibitem{jiang09}
\bibinfo{author}{Jiang, H.}, \bibinfo{author}{Wang, L.}, \bibinfo{author}{Sun,
  Q.-f.} \& \bibinfo{author}{Xie, X.~C.}
\newblock \bibinfo{title}{Numerical study of the topological {Anderson}
  insulator in {HgTe/CdTe} quantum wells}.
\newblock \emph{\bibinfo{journal}{Phys. Rev. B}} \textbf{\bibinfo{volume}{80}},
  \bibinfo{pages}{165316} (\bibinfo{year}{2009}).

\bibitem{anderson58}
\bibinfo{author}{Anderson, P.~W.}
\newblock \bibinfo{title}{Absence of diffusion in certain random lattices}.
\newblock \emph{\bibinfo{journal}{Phys. Rev.}} \textbf{\bibinfo{volume}{109}},
  \bibinfo{pages}{1492} (\bibinfo{year}{1958}).

\bibitem{chen12}
\bibinfo{author}{Chen, L.}, \bibinfo{author}{Liu, Q.}, \bibinfo{author}{Lin,
  X.}, \bibinfo{author}{Zhang, X.} \& \bibinfo{author}{Jiang, X.}
\newblock \bibinfo{title}{Disorder dependence of helical edge states in
  {HgTe/CdTe} quantum wells}.
\newblock \emph{\bibinfo{journal}{New J. Phys.}} \textbf{\bibinfo{volume}{14}},
  \bibinfo{pages}{043028} (\bibinfo{year}{2012}).

\bibitem{girschik15}
\bibinfo{author}{Girschik, A.}, \bibinfo{author}{Libisch, F.} \&
  \bibinfo{author}{Rotter, S.}
\newblock \bibinfo{title}{Percolating states in the topological {Anderson}
  insulator}.
\newblock \emph{\bibinfo{journal}{Phys. Rev. B}} \textbf{\bibinfo{volume}{91}},
  \bibinfo{pages}{214204} (\bibinfo{year}{2015}).

\bibitem{groth09}
\bibinfo{author}{Groth, C.~W.}, \bibinfo{author}{Wimmer, M.},
  \bibinfo{author}{Akhmerov, A.~R.}, \bibinfo{author}{Tworzyd\l{}o, J.} \&
  \bibinfo{author}{Beenakker, C. W.~J.}
\newblock \bibinfo{title}{Theory of the topological {Anderson} insulator}.
\newblock \emph{\bibinfo{journal}{Phys. Rev. Lett.}}
  \textbf{\bibinfo{volume}{103}}, \bibinfo{pages}{196805}
  (\bibinfo{year}{2009}).

\bibitem{prodan11}
\bibinfo{author}{Prodan, E.}
\newblock \bibinfo{title}{Three-dimensional phase diagram of disordered
  {HgTe/CdTe} quantum spin-{Hall} wells}.
\newblock \emph{\bibinfo{journal}{Phys. Rev. B}} \textbf{\bibinfo{volume}{83}},
  \bibinfo{pages}{195119} (\bibinfo{year}{2011}).

\bibitem{xing11}
\bibinfo{author}{Xing, Y.}, \bibinfo{author}{Zhang, L.} \&
  \bibinfo{author}{Wang, J.}
\newblock \bibinfo{title}{Topological {Anderson} insulator phenomena}.
\newblock \emph{\bibinfo{journal}{Phys. Rev. B}} \textbf{\bibinfo{volume}{84}},
  \bibinfo{pages}{035110} (\bibinfo{year}{2011}).

\bibitem{yamakage11}
\bibinfo{author}{Yamakage, A.}, \bibinfo{author}{Nomura, K.} \&
  \bibinfo{author}{Imura, K.-I.}, \bibinfo{author}{Kuramoto, Y.}
\newblock \bibinfo{title}{Disorder-Induced Multiple Transition Involving Z2 Topological Insulator}.
\newblock \emph{\bibinfo{journal}{J. Phys. Soc. Jpn.}} \textbf{\bibinfo{volume}{80}},
  \bibinfo{pages}{053703} (\bibinfo{year}{2011}).

\bibitem{guo10}
\bibinfo{author}{Guo, H.-M.}, \bibinfo{author}{Rosenberg, G.},
  \bibinfo{author}{Refael, G.} \& \bibinfo{author}{Franz, M.}
\newblock \bibinfo{title}{Topological {Anderson} insulator in three
  dimensions}.
\newblock \emph{\bibinfo{journal}{Phys. Rev. Lett.}}
  \textbf{\bibinfo{volume}{105}}, \bibinfo{pages}{216601}
  (\bibinfo{year}{2010}).

\bibitem{fu14}
\bibinfo{author}{Fu, B.}, \bibinfo{author}{Zheng, H.}, \bibinfo{author}{Li,
  Q.}, \bibinfo{author}{Shi, Q.} \& \bibinfo{author}{Yang, J.}
\newblock \bibinfo{title}{Topological phase transition driven by a spatially
  periodic potential}.
\newblock \emph{\bibinfo{journal}{Phys. Rev. B}} \textbf{\bibinfo{volume}{90}},
  \bibinfo{pages}{214502} (\bibinfo{year}{2014}).

\bibitem{garate13}
\bibinfo{author}{Garate, I.}
\newblock \bibinfo{title}{Phonon-induced topological transitions and crossovers
  in {Dirac} materials}.
\newblock \emph{\bibinfo{journal}{Phys. Rev. Lett.}}
  \textbf{\bibinfo{volume}{110}}, \bibinfo{pages}{046402}
  (\bibinfo{year}{2013}).

\bibitem{song12}
\bibinfo{author}{Song, J.}, \bibinfo{author}{Liu, H.}, \bibinfo{author}{Jiang,
  H.}, \bibinfo{author}{Sun, Q.-f.} \& \bibinfo{author}{Xie, X.~C.}
\newblock \bibinfo{title}{Dependence of topological {Anderson} insulator on the
  type of disorder}.
\newblock \emph{\bibinfo{journal}{Phys. Rev. B}} \textbf{\bibinfo{volume}{85}},
  \bibinfo{pages}{195125} (\bibinfo{year}{2012}).

\bibitem{lv13}
\bibinfo{author}{Lv, S.-H.}, \bibinfo{author}{Song, J.} \& \bibinfo{author}{Li,
  Y.-X.}
\newblock \bibinfo{title}{Topological {Anderson} insulator induced by
  inter-cell hopping disorder}.
\newblock \emph{\bibinfo{journal}{J. Appl. Phys.}}
  \textbf{\bibinfo{volume}{114}}, \bibinfo{pages}{183710}
  (\bibinfo{year}{2013}).

\bibitem{aufray10}
\bibinfo{author}{Aufray, B.} \emph{et~al.}
\newblock \bibinfo{title}{Graphene-like silicon nanoribbons on {A}g(110): A
  possible formation of silicene}.
\newblock \emph{\bibinfo{journal}{Appl. Phys. Lett.}}
  \textbf{\bibinfo{volume}{96}}, \bibinfo{pages}{183102}
  (\bibinfo{year}{2010}).

\bibitem{kara12}
\bibinfo{author}{Kara, A.} \emph{et~al.}
\newblock \bibinfo{title}{A review on silicene - new candidate for
  electronics}.
\newblock \emph{\bibinfo{journal}{Surf. Sci. Rep.}}
  \textbf{\bibinfo{volume}{67}}, \bibinfo{pages}{1} (\bibinfo{year}{2012}).

\bibitem{davila14}
\bibinfo{author}{Dávila, M.~E.}, \bibinfo{author}{Xian, L.},
  \bibinfo{author}{Cahangirov, S.}, \bibinfo{author}{Rubio, A.} \&
  \bibinfo{author}{Lay, G.~L.}
\newblock \bibinfo{title}{Germanene: a novel two-dimensional germanium
  allotrope akin to graphene and silicene}.
\newblock \emph{\bibinfo{journal}{New J. Phys.}} \textbf{\bibinfo{volume}{16}},
  \bibinfo{pages}{095002} (\bibinfo{year}{2014}).

\bibitem{zhu15}
\bibinfo{author}{Zhu, F.-f.} \emph{et~al.}
\newblock \bibinfo{title}{Epitaxial growth of two-dimensional stanene}.
\newblock \emph{\bibinfo{journal}{Nat. Mater.}} \textbf{\bibinfo{volume}{14}},
  \bibinfo{pages}{1020} (\bibinfo{year}{2015}).

\bibitem{prodan11b}
\bibinfo{author}{Prodan, E.}
\newblock \bibinfo{title}{Disordered topological insulators: a non-commutative geometry perspective}.
\newblock \emph{\bibinfo{journal}{J. Phys. A: Math. Theor.}} \textbf{\bibinfo{volume}{44}},
  \bibinfo{pages}{113001} (\bibinfo{year}{2011}).

\bibitem{orth13}
\bibinfo{author}{Orth, C.~P.}, \bibinfo{author}{Str\"ubi, G.} \&
  \bibinfo{author}{Schmidt, T.~L.}
\newblock \bibinfo{title}{Point contacts and localization in generic helical
  liquids}.
\newblock \emph{\bibinfo{journal}{Phys. Rev. B}} \textbf{\bibinfo{volume}{88}},
  \bibinfo{pages}{165315} (\bibinfo{year}{2013}).

\bibitem{rod15}
\bibinfo{author}{Rod, A.}, \bibinfo{author}{Schmidt, T.~L.} \&
  \bibinfo{author}{Rachel, S.}
\newblock \bibinfo{title}{Spin texture of generic helical edge states}.
\newblock \emph{\bibinfo{journal}{Phys. Rev. B}} \textbf{\bibinfo{volume}{91}},
  \bibinfo{pages}{245112} (\bibinfo{year}{2015}).

\bibitem{stutzer15}
\bibinfo{author}{St\"utzer, S.} \emph{et~al.}
\newblock \bibinfo{title}{Experimental realization of a topological Anderson insulator}.
\newblock \bibinfo{journal}{Paper presented at CLEO: QELS Fundamental Science 2015, San Jose (CA), United States, 10-–15 May 2015, \texttt{doi:10.1364/CLEO\_QELS.2015.FTh3D.2}}.

\bibitem{weeks11}
\bibinfo{author}{Weeks, C.}, \bibinfo{author}{Hu, J.}, \bibinfo{author}{Alicea,
  J.}, \bibinfo{author}{Franz, M.} \& \bibinfo{author}{Wu, R.}
\newblock \bibinfo{title}{Engineering a robust quantum spin {Hall} state in
  graphene via adatom deposition}.
\newblock \emph{\bibinfo{journal}{Phys. Rev. X}} \textbf{\bibinfo{volume}{1}},
  \bibinfo{pages}{021001} (\bibinfo{year}{2011}).

\bibitem{hua12}
\bibinfo{author}{Jiang, H.}, \bibinfo{author}{Qiao, Z.}, \bibinfo{author}{Liu,
  H.}, \bibinfo{author}{Shi, J.} \& \bibinfo{author}{Niu, Q.}
\newblock \bibinfo{title}{Stabilizing topological phases in graphene via random
  adsorption}.
\newblock \emph{\bibinfo{journal}{Phys. Rev. Lett.}}
  \textbf{\bibinfo{volume}{109}}, \bibinfo{pages}{116803}
  (\bibinfo{year}{2012}).

\bibitem{ando06}
\bibinfo{author}{Ando, T.}
\newblock \bibinfo{title}{Screening effect and impurity scattering in monolayer
  graphene}.
\newblock \emph{\bibinfo{journal}{J. Phys. Soc. Jpn.}}
  \textbf{\bibinfo{volume}{75}}, \bibinfo{pages}{074716}
  (\bibinfo{year}{2006}).

\bibitem{ishigami07}
\bibinfo{author}{Ishigami, M.}, \bibinfo{author}{Chen, J.~H.},
  \bibinfo{author}{Cullen, W.~G.}, \bibinfo{author}{Fuhrer, M.~S.} \&
  \bibinfo{author}{Williams, E.~D.}
\newblock \bibinfo{title}{Atomic structure of graphene on {SiO2}}.
\newblock \emph{\bibinfo{journal}{Nano Letters}} \textbf{\bibinfo{volume}{7}},
  \bibinfo{pages}{1643--1648} (\bibinfo{year}{2007}).

\bibitem{fratini08}
\bibinfo{author}{Fratini, S.} \& \bibinfo{author}{Guinea, F.}
\newblock \bibinfo{title}{Substrate-limited electron dynamics in graphene}.
\newblock \emph{\bibinfo{journal}{Phys. Rev. B}} \textbf{\bibinfo{volume}{77}},
  \bibinfo{pages}{195415} (\bibinfo{year}{2008}).

\bibitem{varlet15}
\bibinfo{author}{{Varlet}, A.} \emph{et~al.}
\newblock \bibinfo{title}{{Tunable Fermi surface topology and Lifshitz
  transition in bilayer graphene}}.
\newblock \emph{\bibinfo{journal}{ArXiv e-prints: 1508.02922}}
  (\bibinfo{year}{2015}).

\bibitem{nevius15}
\bibinfo{author}{{Nevius}, M.~S.} \emph{et~al.}
\newblock \bibinfo{title}{{Semiconducting Graphene from Highly Ordered Substrate Interactions}}.
\newblock \emph{\bibinfo{journal}{Phys. Rev. Lett.}} \textbf{\bibinfo{volume}{115}},
  \bibinfo{pages}{136802} (\bibinfo{year}{2015}).

\bibitem{titum15}
\bibinfo{author}{Titum, P.}, \bibinfo{author}{Lindner, N.~H.},
  \bibinfo{author}{Rechtsman, M.~C.} \& \bibinfo{author}{Refael, G.}
\newblock \bibinfo{title}{Disorder-induced {Floquet} topological insulators}.
\newblock \emph{\bibinfo{journal}{Phys. Rev. Lett.}}
  \textbf{\bibinfo{volume}{114}}, \bibinfo{pages}{056801}
  (\bibinfo{year}{2015}).

\bibitem{yang15}
\bibinfo{author}{Yang, Z.} \emph{et~al.}
\newblock \bibinfo{title}{Topological acoustics}.
\newblock \emph{\bibinfo{journal}{Phys. Rev. Lett.}}
  \textbf{\bibinfo{volume}{114}}, \bibinfo{pages}{114301}
  (\bibinfo{year}{2015}).

\bibitem{jotzu14}
\bibinfo{author}{Jotzu, G.} \emph{et~al.}
\newblock \bibinfo{title}{Experimental realization of the topological {Haldane}
  model with ultracold fermions}.
\newblock \emph{\bibinfo{journal}{Nature}} \textbf{\bibinfo{volume}{515}},
  \bibinfo{pages}{237--240} (\bibinfo{year}{2014}).

\bibitem{groth14}
\bibinfo{author}{Groth, C.~W.}, \bibinfo{author}{Wimmer, M.},
  \bibinfo{author}{Akhmerov, A.~R.} \& \bibinfo{author}{Waintal, X.}
\newblock \bibinfo{title}{Kwant: a software package for quantum transport}.
\newblock \emph{\bibinfo{journal}{New J. Phys.}} \textbf{\bibinfo{volume}{16}},
  \bibinfo{pages}{063065} (\bibinfo{year}{2014}).

\bibitem{bruus04}
\bibinfo{author}{Bruus, H.} \& \bibinfo{author}{Flensberg, K.}
\newblock \emph{\bibinfo{title}{Many-Body Quantum Theory in Condensed Matter
  Physics}} (\bibinfo{publisher}{Oxford University Press},
  \bibinfo{year}{2004}).

\bibitem{xu13}
\bibinfo{author}{Xu, Y.} \emph{et~al.}
\newblock \bibinfo{title}{Large-gap quantum spin {Hall} insulators in tin
  films}.
\newblock \emph{\bibinfo{journal}{Phys. Rev. Lett.}}
  \textbf{\bibinfo{volume}{111}}, \bibinfo{pages}{136804}
  (\bibinfo{year}{2013}).

\bibitem{liu11c}
\bibinfo{author}{Liu, C.-C.}, \bibinfo{author}{Jiang, H.} \&
  \bibinfo{author}{Yao, Y.}
\newblock \bibinfo{title}{Low-energy effective {Hamiltonian} involving
  spin-orbit coupling in silicene and two-dimensional germanium and tin}.
\newblock \emph{\bibinfo{journal}{Phys. Rev. B}} \textbf{\bibinfo{volume}{84}},
  \bibinfo{pages}{195430} (\bibinfo{year}{2011}).

\end{thebibliography}

\end{document}